\documentclass[preprint]{aastex}
\usepackage{emulateapj5}
\usepackage{epsfig}

\begin{document}

\title{A Wind Driven Warping Instability in Accretion Disks 
}

\author{A.~C.\ Quillen
}
\affil{Steward Observatory, The University of Arizona, Tucson, AZ 85721;
aquillen@as.arizona.edu}
\affil{Department of Physics and Astronomy, University of Rochester, Rochester, NY 14627; aquillen@pas.rochester.edu}

\begin{abstract}
A wind passing over a surface may cause an instability in
the surface such as the flapping seen when 
wind blows across a flag or waves when wind blows across water.
We show that when a radially outflowing wind blows across a dense 
thin rotating disk, an initially flat disk is unstable to warping.
When the wind is subsonic, 
the growth rate is dependent on the lift generated by the wind
and the phase lag between the pressure perturbation and the vertical
displacement in the disk caused by drag.
When the wind is supersonic, the grow rate is primarily dependent
on the form drag caused by the surface.
While the radiative warping instability proposed by Pringle is promising
for generating warps near luminous accreting objects, we expect the wind
driven instability introduced here would dominate in objects 
which generate energetic outflows.
\end{abstract}


\section{Introduction}

For a wind blowing over a surface the velocity and pressure
are related by Bernoulli's equation; $P + {1 \over 2} \rho v^2 $ is conserved
along streamlines.  Because the velocity
increases as the wind passes over a protrusion of the surface,
the pressure must decrease.  This causes a force pulling
the higher regions of the surface upwards. 
This force also causes lift on airplane wings.
The Kelvin-Helmholtz instability occurs for the same physical
reason at the boundary of two fluids of different densities moving
respect to one another (e.g., \citealt{chandra}).
A related instability exists
for wind passing over water or for wind passing over fabric
(e.g., \citealt{thwaites,mahon}). 

Near accreting compact objects such as active galactic nuclei,
substantial amounts of kinetic energy can be
present in a wind which may pass near denser colder material in the
outer parts of an accretion disk.
In this paper we consider the possibility
that a wind passing over a dense disk can result in a warp
instability similar to that caused by radiative forces
\citep{pringle,maloney96,maloney97}.
Previous work, \citep{shandl,porter}, 
has considered the torque on a disk which
would be caused by ram pressure or the reaction force from a wind 
but has not explored the possibility that the wind/disk interaction
could lift the disk, and so cause a warping instability.

\section{Perturbative response of a radial wind to a corrugated surface}

We divide the problem into two parts, a diffuse outflowing wind
and a dense infinitely thin disk.  We first compute the effect
of a vertical perturbation or ripple in the disk on the wind.  
The flow has a perturbation in the pressure along the surface
of the disk which pushes on the disk.  We then incorporate this force
into the equations of motion in the disk.
This approach is similar to that used to investigate wind/water wave
interactions or wind/fabric interactions.
We follow the perturbative approach of \citet{chandra}.

We describe a warped surface by a displacement in the direction
normal to its undisturbed plane,
\begin{equation}
\label{warp}
h(r,\theta,t) =   {\rm Re}
\left[ S e^{i\left(m \theta - k_r r- \omega t\right)}\right]
\end{equation}
where $S\ll r$.
We assume a primarily radial flow with velocity 
$\vec{u}  = u_0 \hat{e}_r + \vec{v}$
where $\vec{v}$ is a perturbation, $|\vec{v}| \ll u_0$.
We constrain the velocity so the component normal to the surface
is zero; 
${d h \over d t} =  
\left(
{\partial  \over \partial t} + u_0 {\partial \over \partial r } 
\right) h = v_z$.
This constraint implies that on the surface
\begin{equation}
\label{vz}
v_z = {\rm Re} \left[ -i \left( \omega + u_0 k_r \right) 
S e^{i\left(m \theta - k_r r- \omega t\right)}\right].
\end{equation}

The continuity equation,
${\partial \rho \over \partial t}  + \bigtriangledown \cdot \rho \vec{v} = 0$,
in cylindrical coordinates is
\begin{equation}
{1 \over r} {\partial \over \partial r}(r \rho u_r)  + 
{1 \over r} {\partial \over \partial \theta }(\rho u_\theta )  + 
{\partial \over \partial z}(\rho u_z)  + 
{\partial \rho \over \partial t} =0
\end{equation}
where $\rho$ is the density in the wind.
When the wind is subsonic,
we search for a solution for the velocity and pressure perturbations
in the wind that decays exponentially with increasing distance
from the displaced surface.
When the wind is supersonic, we search for a solution
for the velocity and pressure perturbations 
that vary in phase with distance from the displaced surface.
\begin{eqnarray}
P_1     &=& {\rm Re} \left[ p_1 g(r,\theta,z,t) \right] \\
v_\theta&=&  {\rm Re} \left[ v_{\theta,1} g(r,\theta,z,t) \right] \nonumber \\
v_r     &=&  {\rm Re} \left[ v_{r,1} g(r,\theta,z,t) \right] \nonumber \\
v_z     &=& {\rm Re} \left[ v_{z,1} g(r,\theta,z,t) \right] \nonumber
\end{eqnarray}
where
\begin{eqnarray}
g(r,\theta,z,t)  = & 
e^{i\left(m \theta -k_r r -  \omega t\right) - k_z |z|} &
{\rm  for} ~ M<1  \\
                 = & 
e^{i\left(m \theta -k_r r -  \omega t + k_z |z| \right)} &
{\rm  for} ~ M>1  \nonumber
\end{eqnarray}
and the Mach number, $M \equiv {u_0 \over c_s}$.
The pressure, $P=P_0 + P_1$, $P_1 \ll P_0$ and
we adopt an equation of state $P \propto \rho^\Gamma$ with sound
speed in the wind $c_s$.
For the Kelvin-Helmholtz instability in an incompressible fluid
$k_z$ is directly related to $k_x$ where the one fluid
is moving with respect to the other in the $x$ direction.
For incompressible potential flow solutions the vertical
forms of the variables are solved exactly and they
decay exponentially with height when the flow is subsonic
and vary in phase with height when the flow is supersonic
(e.g., \citealt{shivamoggi}).

In the the tight winding limit ($k_r \gg 1/r$) and to first order 
the continuity equation becomes
\begin{eqnarray}
\label{cont}
{1 \over \Gamma} {p_1 \over P_0} (\omega + u_0 k_r) + k_r v_{r,1}
- {m v_{\theta ,1} \over r} & &  \\
   = & i k_z v_{z,1} {\rm sign}z &{\rm for ~} M<1 \nonumber \\
   = & k_z v_{z,1}   {\rm sign}z &{\rm for ~} M>1.\nonumber 
\end{eqnarray}
Euler's equation to first order in the same coordinate system
\begin{eqnarray}
v_{r,t} + u_0 v_{r,r} + u_{0,r} v_r = - {P_{1,r} \over \rho}  \\
v_{\theta,t} + u_0 v_{\theta,r} + {u_0 v_\theta\over r} = - {P_{1,\theta} \over r \rho} \nonumber \\
v_{z,t} + u_0 v_{z,r}  = - {P_{1,z} \over \rho}.\nonumber
\end{eqnarray}
Again in the tight winding limit
\begin{eqnarray}
\label{eulers} 
(\omega + u_0 k_r )v_{r,1}    =& -k_r {c_s^2 \over \Gamma} {p_1 \over P_0}&
\\ 
(\omega + u_0 k_r )v_{\theta,1} =&   {m \over r} {c_s^2 \over \Gamma} {p_1 \over P_0}& \nonumber \\ 
 (\omega + u_0 k_r )v_{z,1} =& i k_z {c_s^2 \over \Gamma} {p_1 \over P_0}
{\rm sign}z & {\rm for}~ M <1  \nonumber \\
                           =&  k_z {c_s^2 \over \Gamma} {p_1 \over P_0} 
{\rm sign}z & {\rm for}~ M >1. \nonumber 
\end{eqnarray}
From Equations (\ref{cont}) and (\ref{eulers})
we find that 
\begin{equation}
\label{disp1}
(\omega + u_0 k_r)^2 + c_s^2 (\pm k_z^2 - k_r^2 - {m^2\over r^2}) = 0
\end{equation}
where the sign of the $k_z^2$ term is positive for $M<1$ and negative for
$M>1$.
Neglecting the $m^2/r^2$ term in the tight winding limit
\begin{eqnarray}
\label{kz}
k_r^2 -  { 1\over c_s^2} (\omega + u_0 k_r)^2  
      = & {k_z}^2  ~ {\rm for ~} M<1 \\ 
      = & -{k_z}^2 ~ {\rm for ~} M>1 \nonumber
\end{eqnarray}
which relates the vertical exponential scale length or wavenumber
to the radial wavenumber.

\includegraphics[angle=0,width=3in]{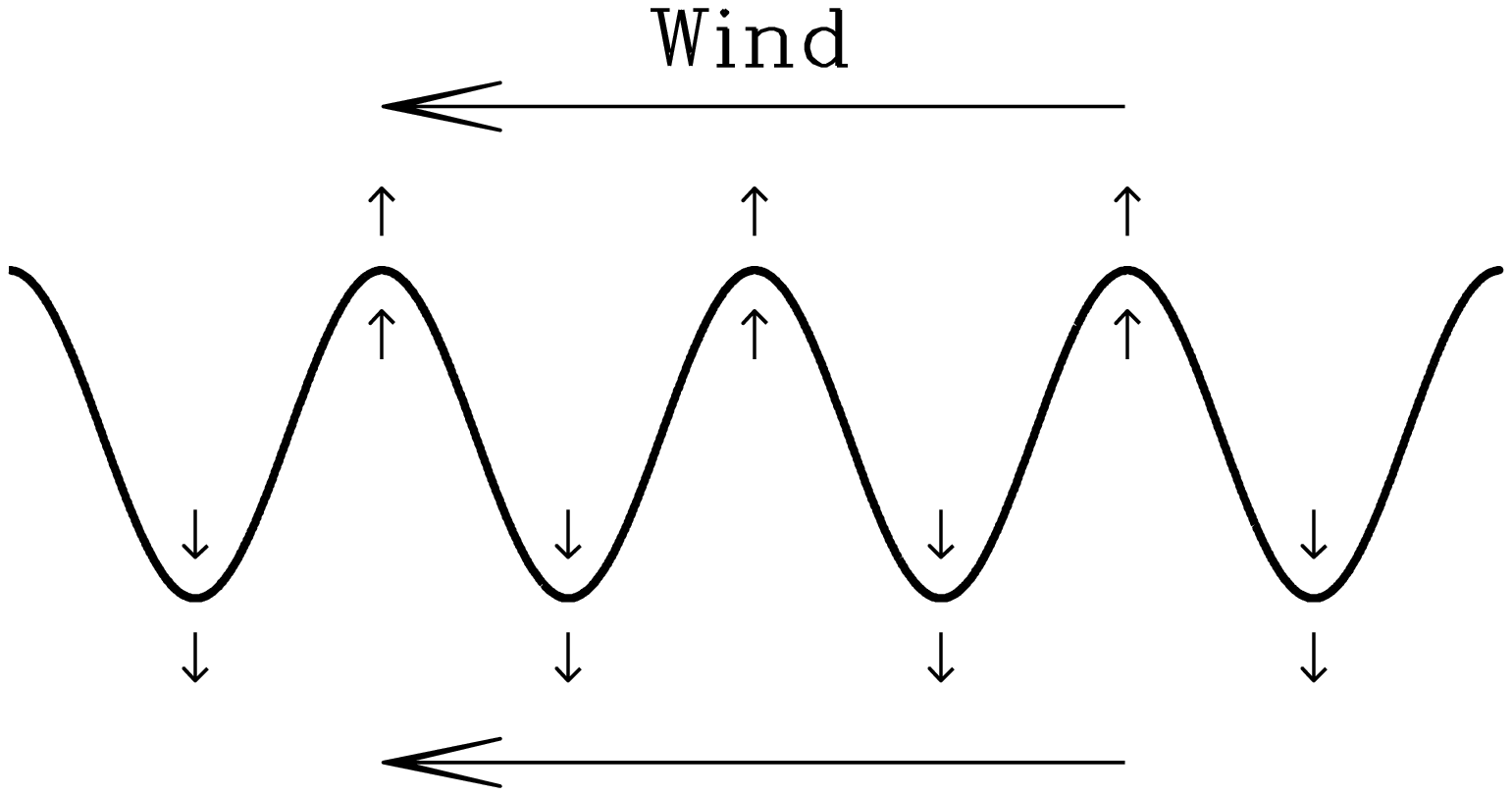}
\begin{quote}
\baselineskip3pt
{\footnotesize Fig.~1-- As a subsonic wind passes over corregations, 
pressure changes cause lift at the high points of the surface.}
\end{quote}
\smallskip

Using Equations(\ref{vz}, \ref{eulers}) we can see that 
\begin{eqnarray}
\label{pressure}
{p_1 \over P_0} = & - {\Gamma \over c_s^2}{ S \over k_z} 
( \omega + u_0 k_r)^2 & {\rm sign}z  ~ {\rm for} ~ M<1 \\ 
                = & - {i \Gamma \over c_s^2}{ S \over k_z} 
( \omega + u_0 k_r)^2 &  ~ ~ ~  ~ ~ ~     ~ {\rm for} ~ M>1  \nonumber 
\end{eqnarray}
When the wind is subsonic, 
the frequencies and wavevectors are real and $k_z>0$, then
$p_1 < 0$ for $z>0$.  Where the surface is high, above the surface
the pressure is lower than the mean and below the surface the pressure 
is higher.  This causes a destabilizing force which we also call 
lift in analogy to the aerodynamics of wings.
Because the sign of the pressure perturbation is opposite
on either size of the disk ($\propto {\rm sign~} z$),
the pressure differential across the surface exerts a force 
on the surface with magnitude twice $P_1$ 
in the direction normal to the surface (see Figure 1).
The pressure differential across the surface caused by the wind
should exert a force on the surface.
An instability should exit when
$m=0$ which is a direct analogy for the Kelvin-Helmholtz instability
or for the instability of a wind passing over fabric.

When the wind is supersonic, the pressure perturbations are 90 degrees
out of phase with the surface perturbations.
The phase of the pressure perturbations should increase with
both $r$ and $|z|$ which implies that the sign of $k_z$ 
is the same as the sign of $k_r$. 
Because the sign of the pressure perturbation is opposite
on either size of the disk, there is a drag force on the surface which
is not present when the flow is subsonic and approximated as a potential
flow solution.   Because shocks should
form we expect the actual flow to be more complicated 
than given by the above equations (see Figure 2 for an illustration).  
When the shocks
and expansion waves lag the high points of
the surface, we expect a vertical
force on the high points of the surface which pushes the surface 
towards the midplane.
When $m=0$, instead of a destabilizing lift force, we expect
a stabilizing vertical force.  But since there is drag,
energy from the wind is transferred to the surface and the
amplitude of the surface perturbation should increase 
(e.g., \S 5.5 of \citealt{shivamoggi} on the Kelvin Helmholtz instability
with a supersonic flow).

\includegraphics[angle=0,width=3in]{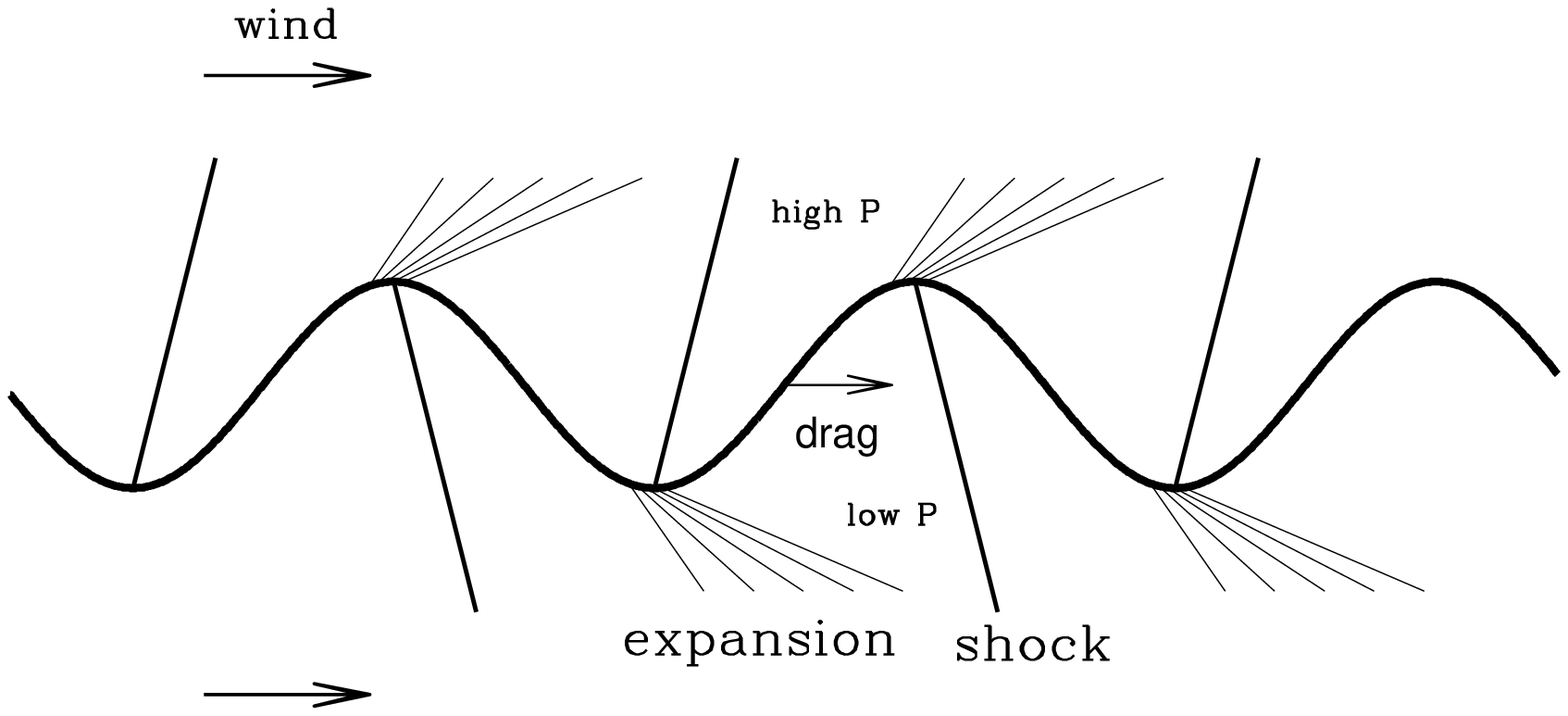}
\begin{quote}
\baselineskip3pt
{\footnotesize Fig.~2-- When the wind is supersonic, the maxima of
the pressure perturbations vary as a function of distance from the surface.
Shocks cause pressure increase, and expansion fans cause
pressure decreases.  The pressure differences across
the surface result in a drag on the surface.
When the expansion waves and shocks lag the high 
points of the surface, there is a vertical force on the surface
which pushes the high points towards the midplane.
}
\end{quote}
\smallskip

When $m=1$ for both subsonic and supersonic flows
we must consider the torque on an annulus,
which is the cross product of the radial vector
and the pressure differential, integrated azimuthally about the ring.

\section{The torque on disk annuli}

To integrate the torque azimuthally we now shift notation and 
follow that used by \citet{pringle} to describe the warped
disk.  The tilt vector for a ring at radius $r$
\begin{equation}
\hat{l} = (\cos{\gamma} \sin{\beta}, \sin{\gamma} \sin{\beta}, \cos{\beta})
\end{equation}
where $\beta(r,t)$ is the local angle of tilt of the disk with respect
to the $z$ axis, and the descending node of the disk material is at an angle 
$\gamma(r,t) - \pi/2$ to the $x$ axis.
The unit vector toward points on the surface
\begin{eqnarray}
\label{xhat}
\hat{x}(r,\phi) &=& (\cos{\phi}\sin{\gamma} + \sin{\phi} \cos{\gamma} \cos{\beta}, \\
& & \sin{\phi} \sin{\gamma} \cos{\beta} - \cos{\phi} \cos{\gamma}, 
-\sin{\phi} \sin{\beta}) \nonumber
\end{eqnarray}
where $\phi=0$ at the descending node and $\gamma$ and $\beta$ 
are both functions of $r$.
The external coordinate system, $\hat{x} = (x,y,z)$,
which in cylindrical coordinates is $(r,\theta,z)$
where $\theta = \tan^{-1}(y/x)$, can be related to that
described by Equation (\ref{xhat}).  When $\beta\ll 1$
\begin{equation}
\theta = \phi + \gamma - \pi/2
\end{equation}
and the vertical displacement of the surface or $z$ component of $\hat{x}$
\begin{equation}
h(r,\theta,t) = -\beta r\sin{\phi} = -\beta r \cos(\theta -\gamma)
\end{equation}
To relate this formalism to that used in the previous section
we set 
$\gamma = k_r r + \omega t$, $m=1$ and $S = -\beta r$ so that
\begin{equation}
h(r,\theta,t) = -\beta r \cos(\theta - k_r r - \omega t)
\end{equation} 
which is in the same form as Equation(\ref{warp}).

\citet{pringle} defines an elemental area vector 
$\vec{dS} = \vec{s_r} \times \vec{s_\phi} dr d\phi$.  
where $\vec{s_r} = {\partial \hat{x} \over \partial r}$
and $\vec{s_\phi} = {\partial \hat{x} \over \partial \phi}$.
The normal to the disk surface
$\hat{n} = {\vec{dS} \over |\vec{dS}|}$.
To first order in $\beta$
\begin{equation}
\vec{s_\phi} = \hat{l} \times \hat{x} = 
\left({
\cos\left(\phi + \gamma\right), \sin \left(\phi + \gamma\right), -\beta \cos{\phi} 
}\right)
\end{equation}
\begin{equation}
\hat{n}  = \hat{l} - \hat{x} (\beta r \gamma' \cos{\phi} - r \beta' \sin{\phi})
\end{equation}
(see \citealt{pringle}, Equations 2.11 and 2.17)
where $\beta'$ and $\gamma'$ refer to the derivatives
of $\beta$ and $\gamma$ with respect to $r$.
%

%

\includegraphics[angle=0,width=3in]{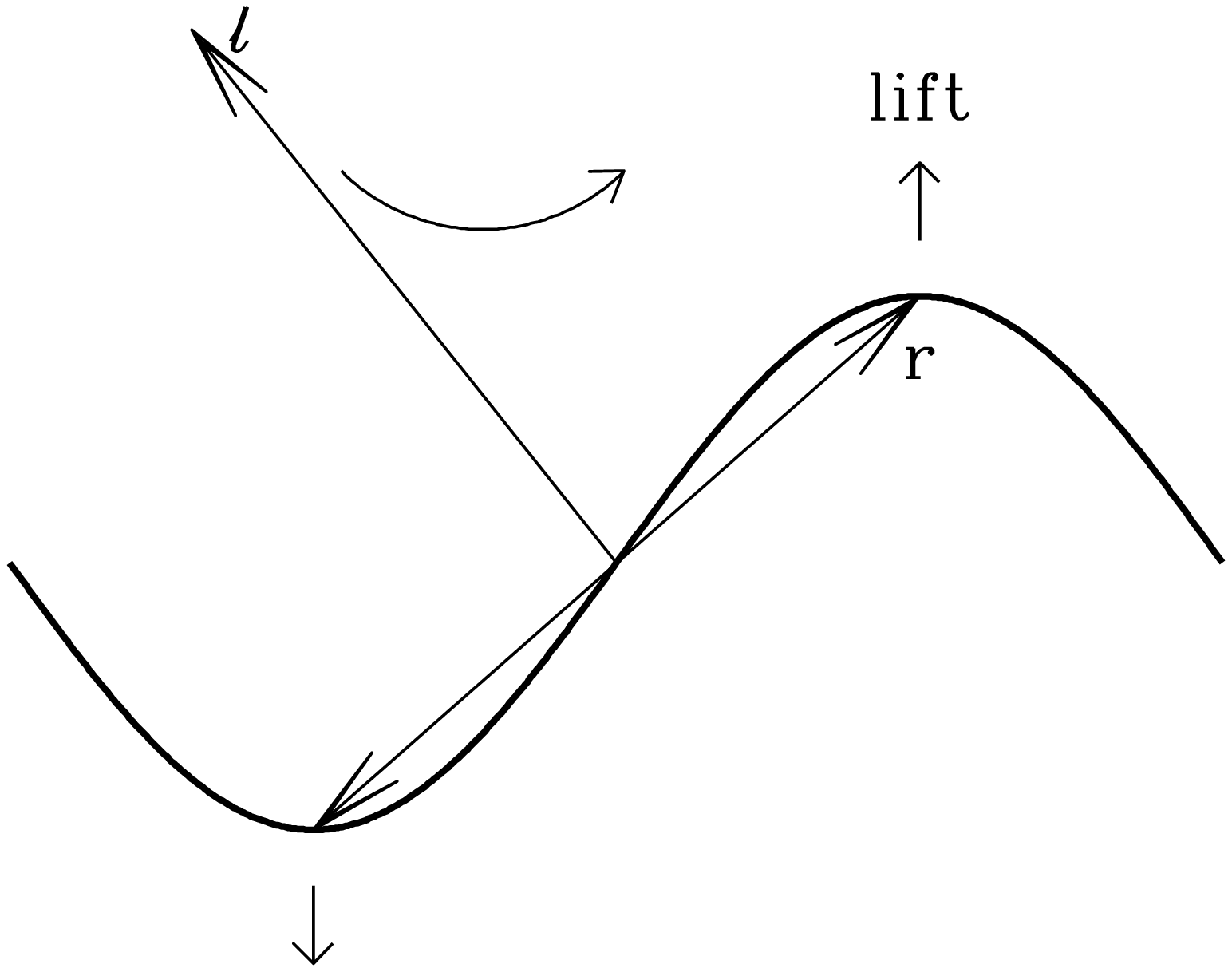}
\begin{quote}
\baselineskip3pt
{\footnotesize Fig.~3-- The lift on an annulus causes the annulus to precess.
When the flow is supersonic, the force on the high points of the surface
is towards the plane and the annulus should precess in the opposite direction.
}
\end{quote}
\smallskip

When the wind is subsonic,
in the previous section we found to first order that 
the pressure differential
across the surface was in phase with the corrugations
of the surface.  However, 
the pressure in the region of laminar flow should actually
be somewhat lower on the leeward
side of each trough and so that there would be a lag between
the pressure and the surface variations \citep{jeffreys}.  
\cite{kendall} measured sinusoidal pressure variations 
in response to a wind passing over a sinusoidally varying ruber surface
and showed that the pressure was indeed offset in phase
with the surface.    
We assume that the pressure difference across
the disk surface can be described 
\begin{equation}
\label{deltap}
\Delta P = 2  p_1  \sin(\phi + \delta)  ~ ~ ~ {\rm for} ~ M<1 
\end{equation}
where the phase lag between the pressure and the surface is given
by $\delta$.   For $k_r > 0$ we expect $\delta > 0$ and
that $\delta<0$ for $k_r <0$.  

When the wind is supersonic, 
shocks should cause an effective phase shift in the pressure perturbation
on the surface.  In this case
\begin{equation}
\label{deltap_supersonic}
\Delta P =  2 p'_1 \cos(\phi + \delta)  ~ ~ ~ {\rm for} ~ M>1 
\end{equation}
where 
\begin{equation}
\label{p1prime}
{p'_1 } \equiv {P_0 \Gamma \over c_s^2} {S \over k_z} (\omega^2 + u_0 k_r)^2 
\end{equation}
see Equation(\ref{pressure}).  
The sign of $p'_1$ depends on
the sign of $k_r$, since the sign of $k_z$ depends on the sign of $k_r$.

The pressure differential across the surface exerts a force
on the disk parallel to the normal to the surface, $\hat{n}$.  
The resulting torque per unit mass
per unit mass for an annulus is the integral  
\begin{equation}
\label{torque_general}
\vec{\tau_w}= {1 \over 2 \pi} \int_0^{2 \pi} r \hat{x} \times \hat{n} 
{\Delta P \over \Sigma} d\phi
\end{equation}
where $\Sigma$ is the disk surface density (mass per unit area).
Using the previous 4 equations, we perform this integral 
for small $\beta$ and $\delta$
\begin{eqnarray}
\label{torque}
\vec{\tau_w}= & {p_1 r\over \Sigma} [
(\sin{\gamma}, -\cos{\gamma}, 0)  - 
\delta (\cos{\gamma}, \sin{\gamma}, 0)   \\ 
& + ( 0,0,\delta \beta)] ~ ~ ~ ~ {\rm for} ~ M< 1 \nonumber \\ 
            = & {p'_1 r\over \Sigma} [
-\delta (\sin{\gamma}, -\cos{\gamma}, 0)   
-  (\cos{\gamma}, \sin{\gamma}, 0)  \nonumber \\ 
& + ( 0,0, \beta) ] ~ ~ ~ ~ ~ {\rm for} ~ M> 1 \nonumber 
\end{eqnarray}
To first order in $\beta$ the angular momentum axis of an annulus
is $\hat{l} = (\beta\cos{\gamma},\beta\sin{\gamma},1)$.
From comparing Equation (\ref{torque}) to $d\hat{l}\over dt$
we can see that the first term in this equation
causes the disk to precess, and the second
term causes the tilt to increase or decrease. 
The tilt increases when $\delta < 0$ or $k_r <0$,  for a subsonic wind
and when $p'_1 <0$ or $k_r < 0$ for a supersonic wind
(see Figures 3 and 4).  
The third term corresponds
to a rate of change in the total angular momentum of the annulus which
would cause a small amount of outflow in the disk.

\includegraphics[angle=0,width=3in]{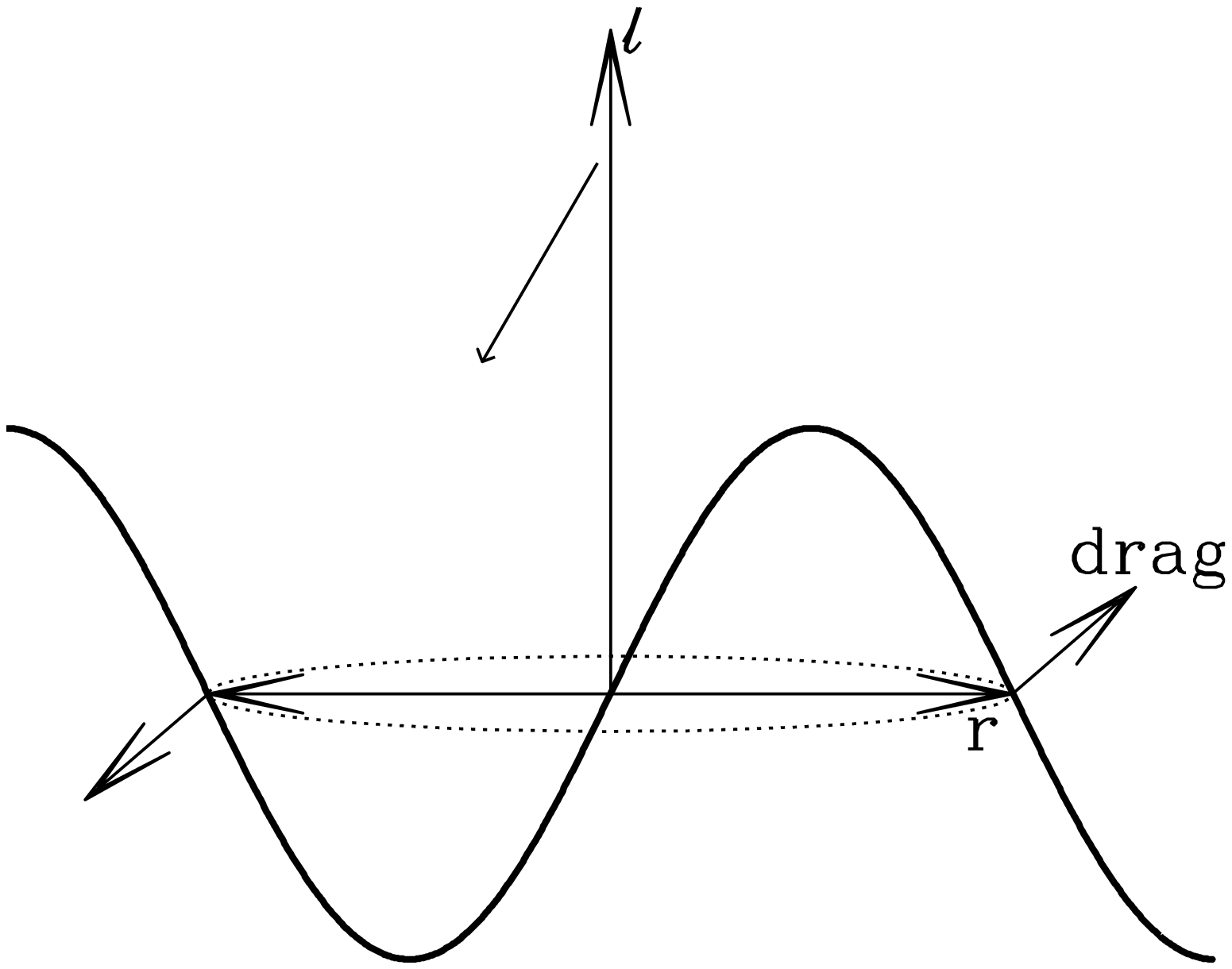}
\begin{quote} 
\baselineskip3pt
{\footnotesize
Fig.~4-- The drag on the surface causes a vertical force on the disk.
This results in a torque which can either increase or decrease
the tilt.   A ring is shown tilted towards the line of sight.
The slope of the warp with respect to the wind depends
upon whether the warp is leading or trailing.  Consequently
the torque either increases or decreases the tilt
depending upon whether the warp is leading or trailing.}
\end{quote}
\smallskip

\subsection{Equation of motion for the annulus}

The angular momentum per unit mass of an annulus of radius
$r$ is $\sim r^2 \Omega$ where
$\Omega$ is the angular rotation rate.
When radial motions in the disk are small
\begin{equation}
r^2 \Omega {\partial \hat{l}\over \partial t} = 
\vec{\tau_g} + \vec{\tau_\nu} + \vec{\tau_w}
\end{equation}
where $\vec{\tau_g}$ is the torque from gravity (equal to zero when
the potential is spherical) and $\vec{\tau_\nu}$ is the torque due
to viscous forces.
As done by \citet{pringle}, we set $W= \beta e^{i\gamma}$
so that $\hat{l} = ({\rm Re}[W],{\rm Im}[W],1)$.
Ignoring the $z$ component of the torque and using
Equation(\ref{torque}) we find 
\begin{eqnarray}
{\partial W \over \partial t} = & -{p_1 \over  \Sigma r \Omega} 
\left[i + \delta \right]{W \over \beta} & ~ ~ {\rm for } ~ M<1 \\
                              = &  {p'_1 \over  \Sigma r \Omega}
\left[1 + \delta i \right]{W \over \beta} & ~ ~ {\rm for } ~ M>1  \nonumber
\end{eqnarray}
where we have ignored gravitation and viscous terms.
Using Equations(\ref{kz}, \ref{pressure}, \ref{p1prime}) for $p_1$
and $p'_1$, 
\begin{eqnarray}
{\partial W \over \partial t}   \approx &
-{ \Gamma P_0 |k_r| M^2 \over \Sigma \Omega \sqrt{1-M^2}}
\left(i + \delta \right) W  & ~ ~ {\rm for} ~ M<1  \\
\approx & { \Gamma P_0  k_r  M^2 \over \Sigma \Omega \sqrt{M^2-1}}
\left(1 + i \delta \right) W  & ~ ~ {\rm for} ~ M>1  \nonumber
\end{eqnarray}
where we have assumed slowly varying modes, $\omega \ll u_0 k_r$.
We define a parameter $Q$ which describes the ratio of the
kinetic energy in the wind to the rotational energy in the disk,
$Q \equiv { P_0 M^2 \over \Sigma \Omega^2 r }
\sim { \rho_w r u_0^2 \over \Sigma v_c^2 }$
where $\rho_w$ is the density of the wind and $v_c \equiv \Omega r$
is the velocity of a particle in a circular orbit.
We can write
\begin{eqnarray}
\label{Wequation}
{\partial W \over \partial t}  \approx  &
-{Q \Gamma |k_r  r| \Omega  \over \sqrt{1-M^2}}
\left(i + \delta \right) W & ~ ~ {\rm for} ~ M<1 \\
                               \approx &
 {Q \Gamma  k_r  r  \Omega  \over \sqrt{M^2 - 1}}
\left(1 + i \delta \right) W & ~ ~ {\rm for} ~ M>1 \nonumber 
\end{eqnarray}
The precession rate 
\begin{eqnarray}
\label{omegaQ}
{\partial \gamma \over \partial t} 
\approx \omega \approx &
-{Q |k_r  r| \Omega}         & {\rm for} ~ M\ll 1 \\
              \approx & 
{\delta Q |k_r  r| \Omega/M} & {\rm for} ~ M\gg 1 \nonumber
\end{eqnarray}
where to simplify the expression we have assumed $M\ll 1$ or $M\gg 1$
and ignored the dependence on the adiabatic index.

For a subsonic wind,
growing modes occur when $\delta<0$.
We expect that the phase lag $\delta<0$ when $k_r <0$ so that
growing modes will occur for $\gamma' < 0$.
The growing mode will have lines of nodes
following a trailing spiral instead
of a leading spiral as was the case for the 
radiatively driven warp instability \citep{pringle}.
For a supersonic wind, growing modes occur when $k_r >0$ and so
will have a leading spiral shape.

We expect the growing mode to grow as $\beta \propto  e^{t/t_{w}}$
with growth timescale 
\begin{eqnarray}
\label{growth_dynamical}
t_w \sim  & {1  \over \delta \omega} & ~ ~ {\rm for} ~ M < 1 \\
    \sim  & {\delta  \over   \omega} & ~ ~ {\rm for} ~ M > 1.\nonumber
\end{eqnarray}
In terms of the dynamical time, $t_d = {1\over\Omega}$,
\begin{eqnarray}
{t_w \over t_d} \sim  & {1  \over Q |k_r r | \delta} ~ ~ {\rm for} ~ M < 1 \\
                \sim  & {M  \over Q |k_r r |       } ~ ~ {\rm for} ~ M > 1. 
\nonumber 
\end{eqnarray}
When the wind is supersonic, the growth rate exceeds the precession rate
and the amplitude of the perturbation can grow quickly.
We see that the precession and growth rates depend on $Q$, the ratio of the 
kinetic energy in the wind and the rotational energy in the disk.
Though we have neglected the viscous forces (in Equation \ref{torque_general}),
we expect them to damp the growth of unstable modes primarily
for the short wavelength modes.
We expect the disk self gravity to affect the precession frequency but not 
strongly affect the growth rate of the mode.

Since we have followed the notation by \citet{pringle}, we can
directly compare the importance of the torque caused
by the wind to that caused by the absorption of radiation from 
a central source. 
Ignoring the viscous damping term, the precession rate
caused by the radiative torque is given by
$\omega_r = {L k_r \over 12 \pi \Sigma R^2 \Omega c}$
(Equations 3.6, 3.8 by \citealt{pringle}).
Comparing this to our Equations(\ref{Wequation},\ref{omegaQ})
we find that the wind driven warping instability is likely
to dominate when
\begin{equation}
{L \over 12 \pi c} \lesssim P_0 M^2 r^2 \sim \rho_w u_0^2 r^2.
\end{equation}
We can interpret this inequality in terms of the momentum flux.
When the momentum flux from the wind dominates that from
a central radiation source, a wind driven instability may
dominate.  This suggests that accreting sources which impart more energy
into driving winds than into radiation would be more likely 
to drive wind driven instabilities in their accretion disks.

\section{The phase lag and the critical angle for boundary
layer separation}

The simplest description of subsonic wind flow near a corrugated surface
such as a flag or water wave
is an irrotational flow above the interface with wind
velocity averaged over one wavelength that
is constant with height.  This is a ``potential flow''
\citep{lamb} and since the pressure is related to the wind velocity
via Bernoulli's equation, it is in exact anti-phase with
the surface (as derived perturbatively above).
So the energy flux between the wind and wave
is zero.  For the wind to do work on the wave, the
pressure must be shifted in phase relative to the potential
solution.  \citet{jeffreys,jeffreys2} suggested that the downwind side
of the wave is sheltered from the wind so that the pressure
on this slope is reduced there but increased on the upwind
side.    This causes a drag force, and as a result
waves traveling with the wind increase in amplitude
and those traveling against the wind are damped.  Scaling based on 
idealized theory for this (\citealt{miles}, and subsequent
work) unfortunately does not predict the experimentaly measured 
growth rates very well (e.g. \citealt{kendall,riley,donelan}). 

In the theory outlined by \citet{miles}, the phase lag
is set by numerical constants determined from turbulent
mean profiles and does not depend upon the amplitude of
the surface variations.  We can understand this 
considering the difference in the rate
that a turbulent boundary layer grows on the leeward and windward
sides of a wavy surface.   Because the growth rate of the boundary
layer depends on the pressure gradient along the direction of flow outside
the boundary layer, we expect
a difference in the boundary layer thickness on the leeward
and windward sides that is proportional to the pressure
difference across the wave, and this in turn is proportional to
the amplitude of the wave.
However, the thickness of the boundary layer may be set by a turbulent
eddy viscosity and so should also be proportional to the amplitude of the
surface perturbation.   Thus, we predict that the phase angle is not be
strongly dependent on the amplitude of the surface variations.

In most astrophysical situations we expect high Reynolds
number flows and so can draw on the theory of aerodynamics.   
A laminar or viscous boundary layer next to a surface
is only stable when the pressure gradient along the surface 
in the direction of flow is negative.
However this condition is violated
to the leeward of a corrugation. 
The same situation occurs along airplane wings and 
the boundary layer becomes turbulent.  
Because of the Coanda effect,
turbulent boundary layers are less likely to separate 
from the surface than viscous
boundary layers.  This allows the boundary layer to
remain near the surface along airplane wings even though
the pressure gradient in the laminar flow region is not
favorable.
For most airplane wings the boundary layer does not separate from
the surface until the angle of attack from the wing is
$\gtrsim 15^\circ$, though this critical angle
can increase to $45^\circ$ for short thick wings.  
Past this critical angle, little or no lift is generated
and the wing `stalls'.  We expect a similar effect
in our astrophysical disk. 
When the amplitude of the corrugations reaches
a critical slope, (a critical value for $|S k_r|$), we expect that the
boundary layer will separate from the surface, the lift
will decrease causing the precession rate to drop, and
the drag will increase.  At this point we expect the mode to saturate.


When the wind is supersonic, a smooth continuous flow is unlikely
(see Figure 2).  A full calculation would require resolving shocks and expansion
flows.  However the perturbative method we have used should allow us
to at least estimate the magnitude of the instability.
We do expect a major difference in the character of the wind flow near
the surface when
the slope of the surface exceeds the Mach angle, or when 
$| S k_r| > {\rm atan} M^{-1}$.  Past this angle a turbulent boundary 
layer should develop and the instability may saturate. 
In this regime it may be possible to estimate the torque on the disk using 
an approximation (that of Newtonian flow) 
which primarily considers ram pressure or reactive force on the disk surface 
(e.g., \citealt{porter,shandl}).

\section{Discussion and Summary}

In this paper we have outlined a possible instability
which could occur in accreting objects with energetic outflows.
A wind passing over a rotating thin dense disk
is likely to cause a warping instability to grow in the disk.
This instability could occur in situations where an energetic
outflow driven by a compact object
passes over a dense disk, such as might happen 
in binary X-ray sources or active galactic nuclei.
In active galactic nuclei, at radii 
of order a parsec from a black hole there is
evidence for the existence of dense warped disks from  
maser observations (e.g., \citealt{herrnstein}).
Large scale outflows are predicted
for a variety of types of accretion flows and seem to 
be an integral part of these flows.
Observational evidence for disk winds is fairly ubiquitous
and seen in both X-ray binaries \citep{brandt,chiang} and in 
AGNs (e.g., \citealt{murray}).

When a subsonic wind passes over a corrugated dense disk,
the corrugations will cause pressure variations along the
surface of the disk.  These pressure variations
cause lift, which cause the annuli in the disk
to precess.  Form drag caused by the corrugated surface
will cause pressure variations in the disk to lag the vertical
displacement of the disk.  This causes a torque on the warped disk
which can increase the amplitude of the perturbation.
We expect this instability to saturate or cease growing at a critical
slope when the lift decreases due to boundary layer separation.

When a supersonic wind passes over a corrugated disk, the disk
primarily experiences form drag.  The drag causes a torque on the warped
disk which can increase the amplitude of the corrugations.
When there is an effective lag between the pressure and 
the vertical velocity of the surface,
there will be a vertical force on the disk
which causes it to precess.
We expect a major change in the nature of the flow when the slope
of the corrugations exceeds the Mach angle. 
While we primarily expect accreting astrophysical objects to drive 
supersonic outflows, 
if a thick outflowing subsonic boundary layer is created,
the subsonic theory outlined here may be appropriate inside this layer.
The instability can dominate the radiative induced warping
instability in sources where the energy released in an outflow
(either sub or supersonic) exceeds that emitted as radiation.

In this paper we have concentrated on a purely hydrodynamic flow.
However we expect the magnetic field to be dynamically
important both in the disk and wind.  
If the wind and disk are magnetized, flux freezing
would make the wind and disk act as mutually impenetrable bodies.
The wind would be less likely to cause hydrodynamical instabilities
that could destroy the disk.  
In this paper we have focused on the affect of a radially outflowing 
wind on the planarity of a disk in the absence of radiation.
Future investigations could develop
a prediction for the phase lag angle for subsonic wind flows 
(a historically difficult hydrodynamics problem), 
and calculate the details of the shocks which are likely
to be present when the wind is supersonic.
In this paper we have restricted our study to radially
outflowing winds, however
astrophysical outflows typically have a nonzero, radius dependent,
azimuthal velocity.  We have also neglected the role
of gravitational, magnetic and viscous forces.
Future work can consider the role of these forces, include
rotating winds and
explore the effects of both radiation and winds on the planarity
of the disk, particularly in the regime where the winds 
themselves are radiatively driven.



\acknowledgments

This work could not have been carried out without helpful
discussions with Eric Blackman, Bruce Bailey, Wayne Hacker,
and Daniel Proga.
We thank the referee, Norm Murray,
for comments which significantly improved this paper.
Support for this work was provided by NASA through grant numbers
GO-07869.01-96A, GO07868.01-96A, and GO07886.01-96A,
from the Space Telescope Institute, which is operated by the Association
of Universities for Research in Astronomy, Incorporated, under NASA
contract NAS5-26555.



{}

\end{document}